\documentclass{acm}

\usepackage{times}  
\usepackage{array, graphics, graphicx, color, url, xspace, multirow, rotating, epsfig, amsmath, amsfonts}
\usepackage{wrapfig}
\usepackage{tikz}
\usepackage{epstopdf,subfig}
\usepackage[format=plain,labelfont=bf,font=small]{caption}
\usepackage{ragged2e,flushend,enumitem,listings}
\usepackage{blindtext}
\usepackage{acronym} 
\usepackage{hyperref, cleveref}
\usepackage{multirow}
\usepackage{svg}
\usepackage[frozencache,cachedir=.]{minted}
\usepackage{appendix}
\usepackage{tabularx}
\usepackage[small,compact]{titlesec}
\usepackage{multicol}

\usepackage{listings}[mathescape]
\usepackage{color}
\usepackage{outlines}
\usepackage{pifont}

\hypersetup{pdfstartview=FitH,pdfpagelayout=SinglePage}

\newcommand\paragraphb[1]{\noindent{\bf{#1}}}
\newcommand\paragraphi[1]{\noindent\emph{#1}}
\newcommand\pb[1]{\paragraphb{#1}}
\renewcommand\pi[1]{\paragraphi{#1}}
\newcommand\pghi[1]{\paragraphi{#1}}

\newcommand{\bi}{\begin{itemize}}
\newcommand{\ei}{\end{itemize}}

\newcommand{\ie}{\emph{i.e.,}\xspace}
\newcommand{\eg}{\emph{e.g.,}\xspace}

\newcommand{\eat}[1]{}
\newcommand{\figspace}{\vspace{-10pt}}

\newcommand{\allnotes}[1]{}
\renewcommand{\allnotes}[1]{\textit{#1}}

\newcommand{\fedlet}{Comvisor\xspace} %
\newcommand{\fedlets}{Comvisors\xspace} %
\newcommand{\fedctl}{cmCTL\xspace} %
\newcommand{\fedctls}{cmCTLs\xspace} %
\newcommand{\fedcore}{cmDAT\xspace} %
\newcommand{\fedcores}{cmDATs\xspace} %
\newcommand{\fedapps}{cmApps\xspace} %
\newcommand{\fedID}{cmID\xspace} %
\newcommand{\fedIDs}{cmIDs\xspace} %

\graphicspath{{./dia/}}
\begin{document}

\title{Comverse: A Federative-by-Design Platform for Community Computing}
\author{Silvery Fu$^1$, Dylan Reimer$^1$, Siyuan Dong$^1$, Yifei Zhu$^2$, Sylvia Ratnasamy$^1$ \\ $^1$UC Berkeley, $^2$UM-SJTU Joint Institute}
\maketitle

\begin{abstract}
Communities, ranging from homes to cities, are a ubiquitous part of our lives. However, there is a lack of adequate support for applications built around these communities. As a result, current applications each need to implement their own notion of communities, making it difficult for both the app developers and the app users (\ie the community admins and members) to create and use these community apps. In this paper, we argue that \emph{communities should be supported at the infrastructure-level} rather than at the app-level. We refer to this approach as the Platform-Managed Community (PMC). We propose Comverse, a platform designed to this end. Comverse is predicated on the principle of federation, allowing autonomous nodes representing community members to voluntarily associate and share data while maintaining control over their data and participation. Through Comverse, we explore the vision of community computing by showcasing its applicability with real-world community apps. 
\end{abstract}

\section{Introduction}
\label{sec:intro}

The notion of community is prevalent in various aspects of our lives, from micro settings such as our homes and labs to macro ones like residential complexes, cities and beyond~\cite{bates2004community}, shaping our social interactions and experiences. The proliferation of mobile and IoT devices, alongside ubiquitous network connectivity~\cite{barela2022connecting,freefi,ccm,scm}, enables exciting new applications centered on these communities. These range from community data hubs~\cite{woo2022communair}, information services~\cite{kamba2009access,woo2022communair}, safety and security monitoring~\cite{brush2013digital,bolt}, to smart villages, campuses, and cities~\cite{deloitte-campus,cisco-campus,somwanshi2016study,yin2015literature}. We refer to these applications as \emph{community apps}.

A community characteristically exhibits two properties: \emph{autonomy} of individual members and \emph{collaboration} among them. In the context of community apps, autonomy means that individual community members retain control and ownership of their devices and data. Collaboration, meanwhile, means that while maintaining their individual autonomy, members cooperate for mutual benefit. This collaboration can involve pooling together data from individual devices to generate community-wide insights or running shared applications serving the entire community.

Today's community apps, however, often fail to fulfill one or both of these properties. First, these apps typically depend on \emph{community members} to either directly upload their data or share access/control of their devices while the apps are operated and owned by the \emph{community admin}, thus forfeiting the autonomy of individual members. For example, in a smart village or building where each household has its own IoT devices capturing data like air quality or security feeds, with today's applications, individual households must either directly upload their data to a centralized system or grant control of their devices to the system, thus losing or hampering their autonomy~\cite{comfy,cisco-building,svs,woo2022communair}. Second, each application implements its own community support, which leads to a partial view of the community on each application.
Specifically, each application maintains its own distinct representation of the community and its associated data, leading to fragmentation and incompatibility issues and thus hindering effective collaboration.
For example, if two households use different IoT apps for environmental monitoring, these two apps will each have their own distinct version of community data. As a result, while the members belong to the same community, the apps cannot easily integrate or cross-reference their data. 

We refer to this current approach as the \emph{app-managed community}. This application-centric approach not only compromises autonomy and hinders collaboration among members, but it also leads to repetitive work for both community app developers---who must build in community management---and app users---who have to manage their community memberships and data across every application. As a result, the scope of today's community apps is limited by the inherent privacy and security concerns as well as the difficulties associated with development and use. They are typically confined to small-scale deployments~\cite{svs,woo2022communair,bates2004community}, experimental use cases~\cite{woo2022communair,brush2013digital}, and research testbeds~\cite{chu2014strack,eagle2009community}.

How can we address these problems? How can we shift the focus to building applications that center around communities, as opposed to constraining communities within the confines of applications? Our insight is that \emph{community should be supported at the infrastructure layer not the application layer}, leading to \emph{platform-managed community} (PMC).\footnote{In this work, we use the terms ``infrastructure'' and ``platform'' interchangeably. Both refer to the underlying software system layer that is shared among and supports the various applications.} 

Specifically, we envision a community platform that provides native support for community and community apps, akin to cloud computing, where a cloud platform provides a variety of cloud services to support cloud applications. However, unlike cloud computing, where the platform is owned by a single entity (\eg the cloud provider), the community platform includes devices and other resources individually owned and managed by each member. Therefore, to preserve the autonomy of members, we argue that \emph{the platform must be federative-by-design} --- platform should be a federation, consisting of components independently operated, rather than being operated and controlled by a single admin. For instance, consider the previous example of two households using separate applications for environmental monitoring. With PMC, these applications would not operate in isolation. Instead, they would interface with a shared platform that federates the entire community's data. So, if household A's app records a sudden increase in temperature, and household B's app detects smoke, the platform would bring and leverage these two pieces of information together.

This paper builds on the above insights and proposes Comverse, a federative-by-design platform that supports platform-managed communities. %
In Comverse, each community and its individual members are represented by a module called a \fedlet, which consists of a control plane component, \fedctl, and a data plane component, \fedcore. At a high level, a \fedlet manages and shares data with community apps and other \fedlets. Specifically, the \fedctl implements community membership tracking, authentication, and authorization, while the \fedcore handles data management and privacy-preserved sharing between members and their community. The \fedlet coordinates its two components, ensuring that the data access occurring at the \fedcore reflects the membership and access control information in the \fedctl. \fedlet exposes a declarative interface for app developers to specify what and how community data in the \fedcore should be handled, as well as runtime APIs for the app logic to interact with the \fedcore. %
The \fedlet can be composed with multiple nested levels (\S\ref{sec:motif}), allowing a community's \fedlet to also serve as another's member.

Comverse introduces three key benefits to community apps with these designs: (i) The \fedlet of each community member serves as an \emph{interposition} between the member and the community admin, enabling members to control when and what data they share. (ii) Now that both the control plane and data plane functionalities are provided at the infrastructure level, collaboration can be achieved through shared protocols between the community and its members' \fedlets. (iii) Apps and app developers can reuse the APIs exposed by the infrastructure, similar to using cloud APIs. Developers and users only need to maintain their membership in one place, rather than in different APIs and different apps. 

The rest of this paper delves into the motivation (\S\ref{sec:motif}), the design proposal of PMC with Comverse (\S\ref{sec:proposal}), followed by an early case study (\S\ref{subsec:study}) and our vision for \emph{community computing} and potential research directions (\S\ref{sec:disc}).

\section{A Case for Platform-Managed Community}
\label{sec:motif}

This section presents a primer on the notion of community and the community apps~\S\ref{subsec:community}, followed by an overview of platform-managed community~\S\ref{subsec:pmc}.

\subsection{What is a Community?}
\label{subsec:community}

Community is a fundamental aspect of human social organization, influencing how we interact, share resources, and collectively address challenges~\cite{tonnies2002community}. A community can be characterized as a group of individuals or entities that come together around shared environments, interests, or purposes. This often occurs in a specific geographical location, such as a village or a campus (focus of this paper), but can also be found online, such as in social networks.

\pb{Community applications.} Community apps refer to applications that are designed around the needs and interactions of a community. They are aimed at addressing shared challenges, improving resource allocation, or enhancing the overall quality of life within the community. Through community apps, the members of the community access the resources and information pooled together and derived at the collective insight at the community level. For example, an app could leverage environmental data from different households to monitor air and water quality~\cite{woo2022communair,verma2015towards}, or use energy consumption data to optimize energy use across households~\cite{birt2012disaggregating,firth2008identifying,nelson2008residential}. Other potential applications include community safety monitoring, health tracking, or even location-aware recommendation systems. These applications harness the collective data generated by the community, while ensuring each participating unit retains control over its data.

\pb{Community app users.} We envision there are two types of users for the community apps. \emph{Community members} are individuals who participate in and contribute to the community, often by sharing data (the focus of this paper), resources, or services. Meanwhile, \emph{community admins} oversee and manage the community's functioning, such as leveraging the shared resources or data from members to run and maintain community apps. It's important to note that community admins can also be members of the communities they administer. For example, a person might be an admin of a neighborhood community that runs safety watch apps (\S\ref{subsec:study}), while also a regular member contributing surveillance data.

\subsection{Platform-Managed Community}
\label{subsec:pmc}

What \emph{layer} is best suited to support community operations? We posit that application-layer solutions are insufficient to tackle this challenge and propose a shift towards the platform layer. This leads us to the concept of a \emph{platform-managed community} (PMC). In our model, it is the platform--rather than isolated applications--that oversees critical functions such as member tracking, data management, sharing, and representing community structure across the community apps.

\pb{Community users benefits.} With the PMC, members can collaborate on supporting community apps without needing to individually install and manage them, while preserving their autonomy and not yielding control of their devices. Admins can conveniently deploy new community apps, leveraging the data provided by the members. Besides, because data can be first processed locally and only deliberately shared, with aggregated insights being shared with the admin, the platform provides a mechanism for enforcing member's privacy and autonomy.

\pb{Community app developers benefits.} Community app developers gain from the modularity and abstraction provided by the community platform. By leveraging the \fedlet's APIs, developers can focus on creating features and functionality for their community apps rather than dealing with community management, membership tracking, and data sharing complexities. The platform can also simplify the testing and debugging processes by providing a uniformed environment for (part of the) community app execution. With the pre-built federation mechanisms, developers can design apps that inherently support collaboration and data sharing across different communities while protecting the autonomy and privacy of the community members.

\section{Comverse: A Proposal for PMC}
\label{sec:proposal}

This section presents the proposed design to achieve PMC with its control plane (\S\ref{subsec:control}) and data plane (\S\ref{subsec:data}), followed by a case study (\S\ref{subsec:study}) on community apps aforementioned.

\subsection{Managing Communities in the Control Plane}
\label{subsec:control}

Comverse’s control plane component is realized as \fedctl, responsible for identity management, community membership, authentication and authorization, joining/leaving. \fedctl’s design is motivated by two primary goals: \textit{establishing membership consensus} and \textit{managing data access}.

\begin{figure}[!ht]
     \centering
     \footnotesize
     \includegraphics[width = 0.40\textwidth]{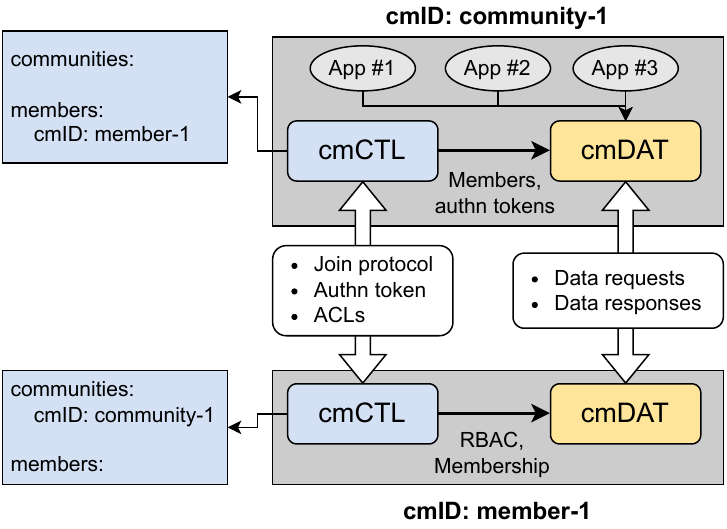}
     \caption{Architecture of \fedlet and control/data federation.}
     \label{fig:fedctl}
\end{figure}

\pb{(1) Identifying member and community.} \fedctl leverages identifiers, \fedIDs, for identity tracking. Each \fedlet is assigned a unique \fedID, which is consistently used throughout the \fedctl design. Rather than creating a new system of identity, \fedctl can use preestablished identities typically present in real-world communities, such as student IDs, employee numbers, or residential unit numbers. This allows the \fedIDs to directly link to real-world identities. For example, a school district could utilize student IDs as \fedIDs in their \fedlet network.

Comverse allows \fedlets to host communities that members can join via the community’s \fedID. Distribution and discovery of \fedIDs can occur in a variety of ways, both through the internet and physically. Communities can advertise their \fedIDs virtually on websites, apps, or online forums, or physically in central community locations. Members can also distribute community \fedIDs to other users to allow them to request to join the community. In order to communicate with a community’s \fedlet, an address for communication must be obtained. One user-friendly option for \fedlet addressing is using a URL naming scheme, which would provide readability and memorability. Encompassed in \fedctl’s approach to identity management is the existence of a public key infrastructure and publicly accessible key storage. This is necessary to authenticate initial communication between communities and members and avoids the need for a three way handshake.

\begin{figure}[!ht]
\centering
\scriptsize
\begin{minipage}{.23\textwidth}
\begin{minted}{yaml}
communities:
  fedID: "u9028599"
    # human-readable name
    name: "Community1"
    # community address
    IP: x.x.x.x
    # community access
    state: active
    # access token
    token: "032njks9w"
    # access permissions
    permitted: {
        "data/temp",
        "data/energy",
    }
\end{minted}
\end{minipage}
\begin{minipage}{.24\textwidth}
\begin{minted}{yaml}
members:
  fedID: "8902340004"
    # human-readable name
    name: "Member1"
    # member address
    IP: x.x.x.x
    # member availablity
    state: active
    # access token
    token: "032njks9w"
    # access permissions
    endpoints: {
        "data/temp",
        "data/energy",
    }
\end{minted}
\end{minipage}
\caption{\textbf{Membership data structures.} Left: Communities this \fedlet has joined. Right: Members of this \fedlet's community.}
\label{code:membership}
\end{figure}

\pb{(2) Tracking membership.} To achieve platform-level federation, \fedctl must be able to facilitate community membership entirely without involvement from \fedapps. Comverse allows nodes to host a community and simultaneously join other communities as a member. Each \fedlet contains two data structures that track relevant community information: the communityList and the membersList, shown in Fig.\ref{code:membership}. The communityList of a \fedlet contains communities that \fedlet has joined, including their \fedIDs, names, addresses, and current data access permissions. Users have the ability to pause or revoke data accessibility from their joined communities. The membersList of a \fedlet contains all current or past members of that \fedlet’s community (if it is hosting a community). The same information in the members struct is contained in the members’ communities structs. This consistency allows synchronicity between communities and their members.

\pb{(3) Joining a community.} Once a community's \fedID and an address for its \fedlet have been obtained, users may send requests to join the community. The community’s \fedlet hosts a web service to receive user communication and send data requests to community members. When a community receives a new join request, it must determine whether to accept or reject the request. The request is first authenticated by verifying its PKI signature, and then the community organizers respond to the request based on community identity. If the community approves the join request, the community’s \fedctl immediately provisions state for the new member. A new entry in the membersList is created with appropriate information, and an approval response is sent to the new member’s \fedlet. If the community denies the join request, no state is needed and a denial response is sent to the user’s \fedlet. Once approved, upon receiving the approval the new member’s \fedctl will likewise provision state for the newly joined community. A new entry in the member’s communityList is created and populated with community information. If denied access to the community, the user may re-request membership at the discretion of the community.

\pghi{(3.1) Authentication.}: \fedctl uses token-based authentication to enforce proper communication between communities and community members. When a new user joins a community, the user’s \fedctl generates a community access token and assigns it to the community’s entry in the communityList. These tokens are generated using unique community information and randomized nonces to eliminate collisions while allowing token regeneration, and are given expiry times. After creation, the access token is sent to the community’s \fedlet and the community stores it inside the users’ corresponding entry in its memberList. To maintain data availability for a community, a member’s \fedctl regenerates and resends the community’s access token on a regular basis. If the community fails to receive an updated token, its \fedctl marks the member as stale and relays this info to relevant \fedapps.

\pghi{(3.2) Authorization.} Once a community member receives and authenticates a data request from a community, an authorization check must occur. \fedctl uses RBAC to create, manage, and enforce access criteria for community members. Upon first joining a community, the new user’s \fedctl creates and stores a role for that community. The user may then create an access list to assign to the community’s role, granting data privileges on a community basis. Modifications to a community’s ACL occur through the \fedctl API as desired by the community member. To enforce this authentication and authorization, \fedctl continuously updates \fedcore with information which we'll discuss in \S\ref{subsec:data}.

After this process, membership consensus has been attained between the community’s \fedlet and the member’s \fedlet. The community is now free to initiate data requests and begin to advertise synchronized data to \fedapps. To leave a community, a community member may send a leave request to the community’s \fedctl. By default, community \fedctls retain information about members that revoke community access and simply modify the member’s state inside the community’s memberList. The member’s updated state is passed to the \fedcore to prevent further access attempts, and by extension relevant \fedapps are notified. Finally, the community concludes by sending an acknowledgement to the user’s \fedlet. 

\begin{table}[t]
\centering
\resizebox{\columnwidth}{!}{%
\begin{tabular}{|l|l|}
\hline
\multicolumn{1}{|c|}{\textbf{API}} & \multicolumn{1}{c|}{\textbf{Functionality}} \\ \hline
comctl list & list community memberships \\ \hline
comctl join \textless{}fedID\textgreater & join a new community \\ \hline
comctl leave \textless{}fedID\textgreater & leave a community \\ \hline
comctl share \textless{}fedID\textgreater {[}data1, …{]} & share datasets with community \\ \hline
\end{tabular}}
\caption{Proposed APIs for \fedctl.}
\label{tab:api}
\vspace{-0.2in}
\end{table}

Community membership operations are executed using the \fedctl APIs, described in Table~\ref{tab:api}. The API may be used directly, or could be used in a more user-friendly application like a phone or web app. Finally, Comverse operates under an assumption that \fedlets are capable of communication via the Internet. \fedlet to \fedlet connectivity is achieved through secure VPN connections~\cite{wiredguard} that are established and maintained by \fedctl.

\subsection{Managing Communities in the Data Plane}
\label{subsec:data}

The \fedcore is responsible for three core tasks: data storage, processing, and communication. In terms of storage, eash \fedcore maintains data pertaining to its own node (community) and the data of its subsidiary nodes (members), which are stored in the form of \textit{objects} and \textit{tables}. When it comes to data processing, a \textit{daemon} embedded within \fedcore executes the data processing logic for the application. The \fedcore daemon exposes an extensive API, facilitating app-specific data configuration within \fedcore and managing data communication with other \fedcores and community apps at runtime. Besides, the \fedcore daemon offers a comprehensive toolkit consisting of utilities such as data encryption, compression, and coordination, all aimed at easing the development process for application providers.

\pb{Tracking and storing data.} The data storage function of \fedcore is critical in achieving \textit{autonomy} for individual members. All raw data is first sent to \fedcore before being accessed by community applications. This means that \fedcore is responsible for storing raw data, as well as aggregated data that the applications can access. This data is stored as \textit{objects} and \textit{tables}, with objects representing a set of \textit{key-value} pairs and tables representing structured data defined by a schema.

Private data from members can be combined into a \textit{materialized view} or \emph{aggregated data} (\eg in the federated learning case, \S\ref{subsec:study}) by the community to provide necessary information to community applications without directly accessing the raw data or compromising privacy. Applications need to specify the purpose, selection and filtering criteria, transformation operations, and the structure of the table. \fedcore daemon handles the updating of these materialized views and notifies the community application.

\pb{Processing data.} \fedcore daemon handles the part of the application logic to assist with \textit{autonomy} and to simplify application development. The daemon preprocesses members' raw data before it is sent to the application and provides common data processing algorithms for reuse by multiple applications. \fedcore daemon provides APIs for the application to dictate the required data processing logic, which is then executed by the daemon.

\pb{Coordination between \fedctl and \fedcore.} When a user joins a community, \fedctl generates an access token, which is then delivered to \fedcore. This token is essential for \fedcore to access and process the user's data. Besides, \fedcore consistently reports the status of data availability and any changes to \fedctl, helping manage permissions and track data lifecycle effectively.

\pb{Coordination across \fedcores.} \fedcores communicate and synchronize data across different nodes. When data changes occur in one \fedcore, it reports the updated data availability to other \fedcores. Moreover, when data from multiple nodes is required, the relevant \fedcores collaborate to consolidate the necessary data, following the data access permissions guided by \fedctl. This coordination ensures timely and efficient data sharing across communities.

These abstractions cover two main areas: management of community data within \fedcore, and the interaction between the application and \fedcore at runtime. The community app needs to provide a file, using the declarative API, that specifies the requirements for data storage, the application state update policy within \fedcore, and the daemon-provided algorithms. This file is then delivered to \fedcore daemon through a programming API, described more next.

\subsection{Development and Deployment}
\label{subsec:dev}

The development process for an application in the Comverse framework can be split into two main components. The first part involves creating the application logic. This comprises the specific rules, procedures, and operations that define how an application operates and performs its intended functions. The second part involves creating a specification file that interfaces with the \fedcore daemon. This file outlines how the application's states are maintained and synchronized across different nodes.

Applications, represented as $A_p$ (parent node) and $A_c$ (child node), may originate from the same or different providers. Applications are inherently dynamic, allowing them to evolve and improve through multiple versions. The compatibility of versions between $A_p$ and $A_c$ is managed either by an external community organization or through an API accessible to $A_p$. Container versioning tools~\cite{container-versioning} are used to verify compatibility between application versions.

If an application requires a table of member data, this requirement can be defined in the specification file provided to the \fedcore daemon. Based on this file, the \fedcore daemon generates and regularly updates the required data table, accumulating member data at the intervals specified in the file. The application is synchronized with each update. The \fedcore daemon provides a toolkit that includes data processing tools such as encryption algorithms (\eg Homomorphic encryption~\cite{morris2013analysis}, Differential privacy~\cite{dwork2006differential}) and compression algorithms (\eg top-$K$ sparsification~\cite{lin2017deep}, FetchSGD~\cite{rothchild2020fetchsgd}, MinMax~\cite{zhao2022minmax}). Developers need to specify which algorithm is needed for which set of objects, removing the need to implement the algorithms themselves.

\pb{Deploying \fedlet.} All community users, whether administrators or members, can opt to run their \fedlet. They can deploy their \fedlet on local machines and servers (such as Raspberry Pi~\cite{pi2015raspberry}, NUC devices~\cite{cohen2014intel}), in the cloud, or through a \fedlet provider/SaaS. The \fedlet should be capable of processing data from member \fedlets and should also support a comprehensive set of integration tools for ingesting data from various devices and their providers.

\pb{Hierarchical federation.} The design of \fedlets in the control plane and data plane naturally enable nested federation across communities. In the control plane, the design of \fedctl allows a \fedlet to manage its community while also being a member of another community. By joining another community, a \fedlet acquires a new \fedID for that community, interacting with it just like any other member. This functionality empowers a \fedlet to extend its operations into other communities, enabling hierarchical or networked community structures. On the data plane, the design of \fedcore allows a \fedlet to manage data for its community and for the communities it has joined. The \fedlet can contribute its community's data to other communities as per the defined data access control policies.

\section{An Initial Case Study}
\label{subsec:study}

We now explore how Comverse's design may help implement a representative community app, community safety watch, which is often sought after in residential complexes.

\pb{Community safety watch (CSW).} The CSW application~\cite{brush2013digital} uses cameras from community households to detect suspicious activities like crashes or robberies~\cite{brush2013digital,chu2014strack,bolt} in real-time, notifying authorities when needed. The detection model is trained based on data feed from the community members. However, this raises data privacy concerns, as cameras capture private community information. 

\pghi{(a) Developing CSW today.} The CSW applications~\cite{bolt} train some models on distinct image feeds from multiple sources. These image feeds are key frames captured by cameras contributing to the application. CSW performs centralized model training by requesting participating devices to send captured images to an application server for acquisition continuously. This server can then use the aggregated data to update the model stored in the server.

\begin{figure}[!ht]
     \centering
     \footnotesize
     \includegraphics[width = 0.43\textwidth]{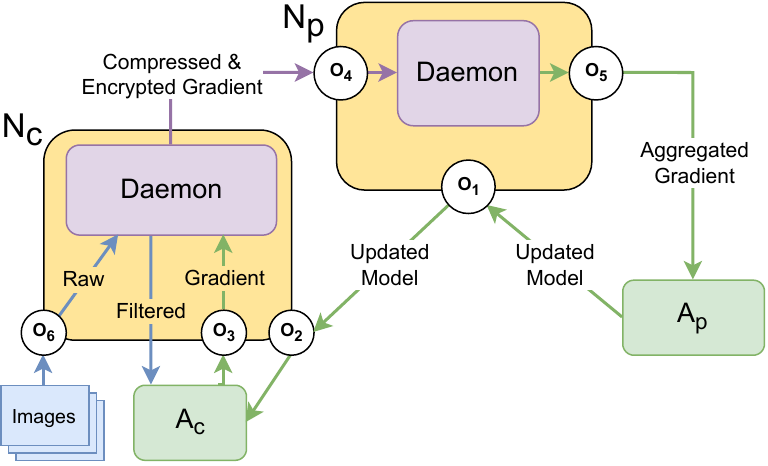}
     \caption{Developing CSW with Comverse.}
     \label{fig:fed_with_comverse}
\end{figure}

\pghi{(b) Developing CSW with Comverse.} Comverse can leverage Federated Learning~\cite{rieke2020future,bonawitz2019towards}, where the CSW application distributes the global model to each device's \fedcore, which performs local training on locally stored data. The \fedcore then calculates the model updates (gradients) and sends these back to the parent node's \fedcore, where they are securely aggregated to update the global model. Fig.~\ref{fig:fed_with_comverse} presents a federated CSW application using Comverse. The parent node $N_p$, the application $A_p$ serving the parent node, the child nodes $N_c$, and the applications $A_c$ serving the child nodes all collaborate in this process. $N_p$'s \fedcore stores an object $O_1$ of the model that will be distributed to every $N_c$, an object $O_4$ to aggregate (sum up) the compressed and encrypted local gradients from the child nodes, and an object $O_5$ to store decrypted aggregated gradients. Each $N_c$'s \fedcore establishes an object $O_2$ that syncs with $O_1$ in the $N_p$'s \fedcore, and an object $O_3$ for storing the local gradient. Each $A_p$ syncs with $O_5$ to monitor the new gradient, and each $A_c$ syncs with $O_2$ to monitor changes in the model. During the training process, key frames (images) captured by the cameras are continuously sent to $O_6$ of $N_c$. The filtered images are sent to $A_c$ for service provision.

\pb{Analysis.} Comverse introduces three key improvements in CSW. First, it preprocesses raw images before they're accessed by the application, preserving autonomy and privacy. Second, it offloads operations such as compression and encryption to the \fedcore daemon, further supporting autonomy and fostering collaboration. Finally, direct data communications are kept strictly between a \fedcore and an app, providing an additional layer of isolation. These design changes lead to two key benefits:

\pghi{(1) Preserving autonomy and privacy.} Comverse allows raw images to be preprocessed before they're acquired by the application, unlike the traditional approach where raw images are sent directly to the application. Further, encryption, aggregation, and decryption processes are performed within Comverse, separating applications from potentially sensitive data operations. Since applications ($A_p$ and $A_c$) can come from different providers, this separation makes collusion more difficult (which establishes individual trust domains for each member).

\begin{table}[]
\centering
\begin{tabular}{|l|l|l|l|}
\hline
\textbf{Function} & \textbf{Module} & \textbf{SLOC} & \textbf{Ratio} \\ \hline
Training & Application & $1205$ lines & $39.42\%$ \\ \hline
\textbf{Federation} & \fedcore & $608$ lines & \textbf{$19.92\%$} \\ \hline
\textbf{Compression} & \fedcore daemon & $913$ lines & \textbf{$29.55\%$} \\ \hline
\textbf{Encryption} & \fedcore daemon & $338$ lines & \textbf{$11.11\%$} \\ \hline
\end{tabular}
\caption{Potential development efforts saving with Comverse for CSW.}
\label{tab:fed_learning}
\vspace{-0.2in}
\end{table}

\pghi{(2) Simplifying community app development.} Comverse  alleviates the burden of implementing federated learning and data processing functionalities. Developers no longer need to build federation capabilities, manage data compression, and coordinate learning across community members, all of which can make up nearly 60.58\% of development efforts (19.92\% for federation, 29.55\% for compression, and 11.11\% for encryption according to Table~\ref{tab:fed_learning}, estimated using an open-source, federated learning-based image processing library~\cite{kvsagg}). Instead, the application's role becomes simplified: $A_p$ receives the gradient from \fedcore and updates the model, while $A_c$ only needs to retrieve the model from \fedcore and generate the gradient. As a result, Comverse distills application tasks to standard machine learning operations, potentially reducing the lines of code written for these functionalities by over half, making the development process significantly more manageable.

\section{Discussion and Conclusion}
\label{sec:disc}

\begin{table}[t]
\def\arraystretch{1.2}
\centering
\footnotesize
\begin{tabular}{|l|l|}
\hline
\textbf{Cloud Computing} & \textbf{Community Computing} \\ \hline
Datacenter & Community \\ \hline
Public Cloud & Federation of Communities \\ \hline
Cloud Provider & Community Admin \\ \hline
Hypervisor/VM & Comvisor \\ \hline
Cloud Service & Community App \\ \hline
Multi-tenant Architecture & Federative Architecture \\ \hline
\end{tabular}
\caption{Analogy between cloud computing and community computing.}
\label{tab:parallel}
\figspace
\end{table}

In this paper we propose Comverse, a federative-by-design platform, aimed at redefining how community applications are developed and used. This shift to a platform approach leads us to the idea of \pb{community computing}, a new way of thinking about and designing applications that are based on platform-managed communities. As shown in Table~\ref{tab:parallel}, we can draw certain parallels and compare the well-understood cloud computing with community computing. For example, datacenters are analogous to individual communities in community computing, while public clouds, comprised of multiple datacenters under one provider, are akin to federations of members/communities. The role of the cloud provider is fulfilled by the community admin in the community computing model, where hypervisor that manage the compute resources in a cloud; Comvisors fulfill this role in the community computing. With this vision in mind, we raise the following research questions: 

\pghi{\bf How applicable is the PMC approach?} The applicability of Comverse and the PMC approach to various community scenarios and application requirements is an open area for exploration. Future research could aim to understand how PMC can be extended to support diverse community settings such as online communities~\cite{discord} and social networks~\cite{facebook}.

\pghi{\bf How performant and scalability is Comverse?} Evaluating Comverse's scalability and understanding the performance implications of scaling are our future work. Key questions to explore include: How does system performance evolve as the number of members increase? How effectively can Comverse handle large-scale data management and processing? Further, it is also important to understand Comverse's tolerance and resilience against failures, such as network disconnections between the \fedlets.

\pghi{\bf What is the potential for convergence of community infrastructure?} An interesting research direction is to explore the co-design and co-deployment of community networking~\cite{ccm,scm} with the data and service infrastructure, as exemplified by Comverse. Community networking serves as the essential connectivity fabric among community members (analogous cloud networking in cloud computing), while the data and service infrastructure manages and provides insights and supports community apps. Exploring the interplay between these two domains could lead to a more integrated and efficient community computing environment.

\onecolumn \begin{multicols}{2}
\bibliographystyle{abbrv} 
\bibliography{comverse}
\end{multicols}

\end{document}